\documentclass[aps, pra, reprint, superscriptaddress, showkeys, nofootinbib]{revtex4-1}

\usepackage[english]{babel}			
\usepackage[utf8]{inputenc}			
\usepackage[T1]{fontenc}			
\usepackage{microtype}				

\usepackage{color} 						
\definecolor{red}{rgb}{1,0,0}					
\definecolor{blue}{rgb}{0,0,1}					
\definecolor{black}{rgb}{0,0,0}				

\usepackage{soul} 
\definecolor{hlyellow}{rgb}{0.95,0.95,0}
\sethlcolor{hlyellow}
\soulregister \ref 7
\soulregister \cite 7
\soulregister \citet 7
\soulregister \footnote 8
\soulregister {\ } 7
\soulregister \figref 7
\soulregister \tabref 7
\soulregister \secref 7

\bibpunct{[}{]}{,}{n}{,}{,} 
\usepackage{natmove} 

\usepackage{amsmath,amssymb,amsfonts}	
\usepackage{array} 

\usepackage{graphicx}			

\usepackage{hyperref}  
\definecolor{dullmagenta}{rgb}{0.4,0,0.4}   
\definecolor{darkblue}{rgb}{0,0,0.4}
\definecolor{medblue}{rgb}{0,0,0.7}
\definecolor{lightblue}{rgb}{0,0,0.8}
\hypersetup{
	colorlinks=true, 		
	breaklinks=true,		
	linkcolor=lightblue,
	citecolor=lightblue,
	filecolor=dullmagenta,
	urlcolor=lightblue
}
\usepackage[all]{hypcap} 	

\renewcommand{\thesection}{\arabic{section}}

\makeatletter 
\renewcommand{\p@section}{}
\renewcommand{\p@subsection}{}
\renewcommand{\p@subsubsection}{}
\makeatother

\usepackage{physmaths}
\usepackage{physunits}

\newcommand{\etal}{\textit{et al.}}

\newcommand{\figref}[1]{Fig.\ \ref{#1}} 
\newcommand{\tabref}[1]{Tab.\ \ref{#1}} 
\newcommand{\secref}[1]{Sec.\ \ref{#1}} 
\newcommand{\figletter}[2]{\mbox{(\textbf{#1})}~#2} 
\renewcommand{\paragraph}[1]{\textbf{#1. }}

\addto\captionsenglish{} 
\addto\captionsenglish{} 
\newcommand{\figtitle}[1]{#1 \normalfont}

\newcommand{\RefSupSecFabAndMeasurement}{Supplementary \secref{sec:fabmeas}}
\newcommand{\RefSupSecFewElectrons}{Supplementary \secref{sec:fewel}}
\newcommand{\RefSupSecEffectiveTwoZero}{Supplementary \secref{sec:effectivedouble}}
\newcommand{\RefSupSecReadout}{Supplementary \secref{sec:spinprepread}}
\newcommand{\RefSupSecPulseAndCali}{Supplementary \secref{sec:pulseseq} and \ref{sec:probcal}} 
\newcommand{\RefSupSecModel}{Supplementary \secref{sec:stmodel}} 
\newcommand{\RefSupSecVisibility}{Supplementary \secref{sec:factorsvisi}} 
\newcommand{\RefSupSecCoherenceTime}{Supplementary \secref{sec:T2analysis}} 
\newcommand{\RefSupSecNoise}{Supplementary \secref{sec:detnoisemodel}} 

\begin{document}

\newcommand{\mytitle}
{Coherent coupling between a quantum dot and a donor in silicon}
\newcommand{\mytitlenobreak}
{Coherent coupling between a quantum dot and a donor in silicon}
\title{\mytitle}

\author{Patrick \surname{Harvey-Collard}}
\email[Corresponding author: ]{P.Collard@USherbrooke.ca}
\affiliation{Département de physique et Institut quantique, Université de Sherbrooke, Sherbrooke, QC, J1K~2R1, Canada}
\affiliation{Sandia National Laboratories, Albuquerque, NM, 87185, United States}

\author{N. Tobias \surname{Jacobson}}
\affiliation{Center for Computing Research, Sandia National Laboratories, Albuquerque, NM, 87185, United States}

\author{Martin \surname{Rudolph}}
\affiliation{Sandia National Laboratories, Albuquerque, NM, 87185, United States}

\author{Jason \surname{Dominguez}}
\affiliation{Sandia National Laboratories, Albuquerque, NM, 87185, United States}
\author{Gregory A. \surname{Ten Eyck}}
\affiliation{Sandia National Laboratories, Albuquerque, NM, 87185, United States}
\author{Joel R. \surname{Wendt}}
\affiliation{Sandia National Laboratories, Albuquerque, NM, 87185, United States}
\author{Tammy \surname{Pluym}}
\affiliation{Sandia National Laboratories, Albuquerque, NM, 87185, United States}

\author{John King \surname{Gamble}}
\affiliation{Center for Computing Research, Sandia National Laboratories, Albuquerque, NM, 87185, United States}

\author{Michael P. \surname{Lilly}}
\affiliation{Center for Integrated Nanotechnologies, Sandia National Laboratories, Albuquerque, NM, 87185, United States}

\author{Michel \surname{Pioro-Ladrière}}
\affiliation{Département de physique et Institut quantique, Université de Sherbrooke, Sherbrooke, QC, J1K~2R1, Canada}
\affiliation{Quantum Information Science Program, Canadian Institute for Advanced Research, Toronto, ON, M5G~1Z8, Canada}

\author{Malcolm S. \surname{Carroll}}
\email[Corresponding author: ]{mscarro@sandia.gov}
\affiliation{Sandia National Laboratories, Albuquerque, NM, 87185, United States}

\date{October 18$^\text{th}$, 2017}

\begin{abstract}
Individual donors in silicon chips are used as quantum bits with extremely low error rates. However, physical realizations have been limited to one donor because their atomic size causes fabrication challenges. Quantum dot qubits, in contrast, are highly adjustable using electrical gate voltages. This adjustability could be leveraged to deterministically couple donors to quantum dots in arrays of qubits. In this work, we demonstrate the coherent interaction of a \ce{^31P} donor electron with the electron of a metal-oxide-semiconductor quantum dot. We form a logical qubit encoded in the spin singlet and triplet states of the two-electron system. We show that the donor nuclear spin drives coherent rotations between the electronic qubit states through the contact hyperfine interaction. This provides every key element for compact two-electron spin qubits requiring only a single dot and no additional magnetic field gradients, as well as a means to interact with the nuclear spin qubit.
\end{abstract}

\maketitle



The silicon industry’s fabrication capability promises to be a differentiating accelerator for the future development of quantum computers built with silicon quantum bits (qubits). Silicon is, furthermore, an appealing material for qubits because it provides an ultra low decoherence environment \cite{tyryshkin2012}. In particular, extremely high fidelities have been demonstrated for both the electron \cite{witzel2010,tyryshkin2012,zwanenburg2013, muhonen2014b,tracy2016} and nuclear spins \cite{pla2013a} of a single dopant atom in isotopically-enriched silicon nanostructures \cite{muhonen2014b}. Assembling these exceptional solid-state qubits into a full quantum processor, as first envisioned by Kane \cite{kane1998}, will require coupling donor atoms to one another in a controllable way. This has proven extremely challenging, demanding near-atomic precision in the placement of the donors \cite{bielejec2010,weber2014,dehollain2014a,singh2016,gamble2015}. In contrast, single electron spins confined in quantum dots (QDs) \cite{veldhorst2015a, eng2015,kim2014c, takeda2016a} are routinely coupled to one another since quantum dots are highly tunable and fabricated in engineered locations, allowing for controllable and scalable two-qubit interactions \cite{nowack2011,brunner2011,shulman2012a,veldhorst2015a,nichol2017}. For this reason, QDs have been theoretically discussed as intermediates to couple donor qubits \cite{kane1998,kane2000,skinner2003,srinivasa2015,pica2016,tosi2017}. Recently, spin blockade has been observed in a silicon QD-donor device \cite{urdampilleta2015a}. However, the coherent spin coupling between donor- and quantum dot-based qubits has remained elusive. It is the cornerstone advance necessary for exploiting the advantages of these two complementary qubit systems.

\begin{figure*}
   \centering
   \includegraphics{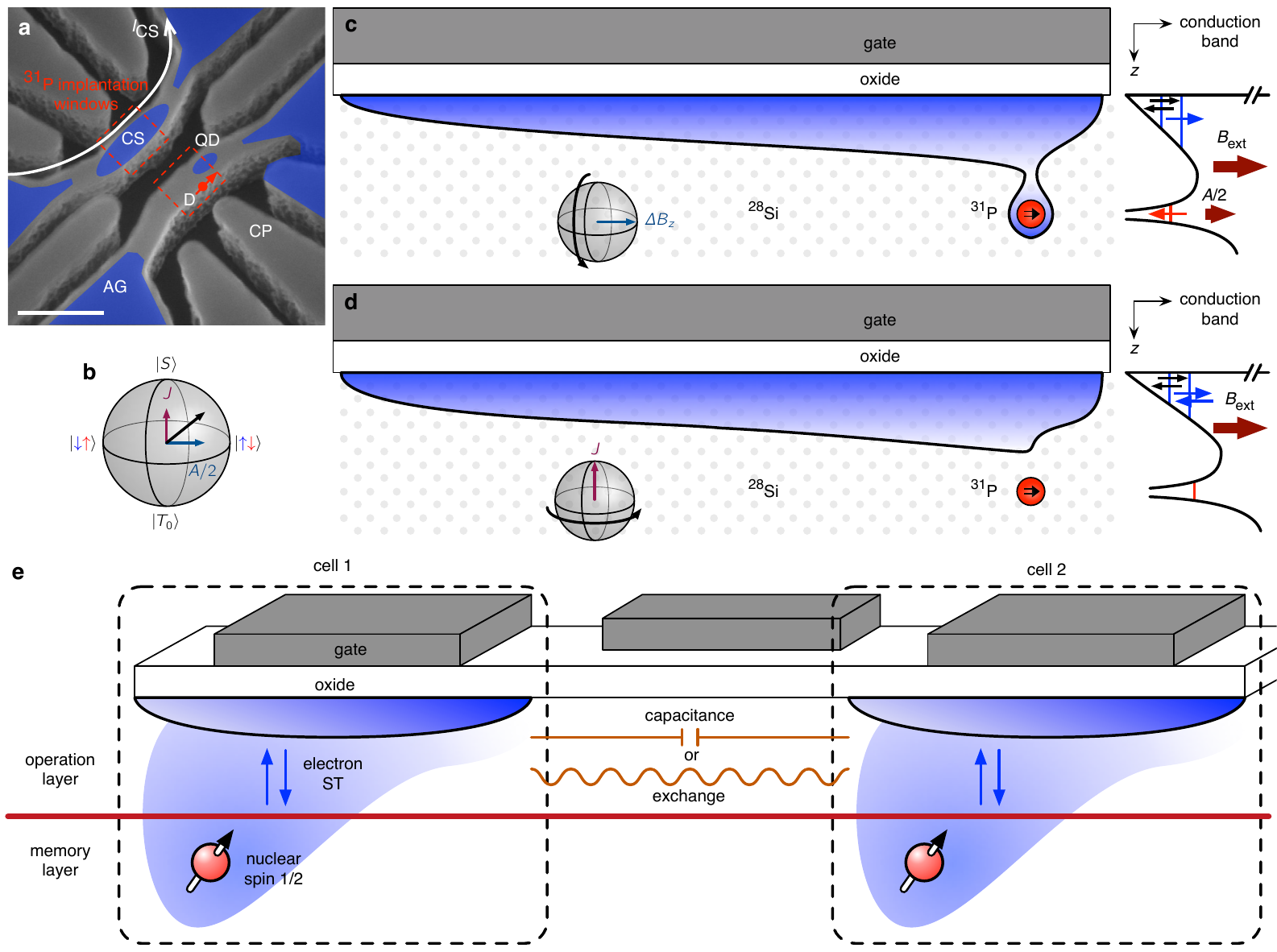} 
   \caption{\figtitle{Quantum dot-donor system.}     \figletter{a} Angled-view scanning electron microscope image of the device gate structure. The blue overlay represents the 2D electron gas at the Si--oxide interface. Donors are implanted in the regions designated by the dashed red lines. The relevant donor (D) located next to the quantum dot (QD) is indicated by the red dot. Scale bar: $200 \nm$.       \figletter{b} A four-electron filled-shell configuration is used to mimic a two-electron singlet-triplet qubit (see main text for details). The Bloch sphere shows the logical singlet $\ket{S}=\ket{\uparrow\downarrow}-\ket{\downarrow\uparrow}$ and triplet $\ket{T_0}=\ket{\uparrow\downarrow}+\ket{\downarrow\uparrow}$ qubit states at the poles. Electrical voltages adjust the relative magnitudes of $J$ and $A/2$: $J$ is dominant in the $(4,0)$ charge configuration, while $A/2$ is in the $(3,1)$ configuration.      \figletter{c} Schematic showing the electrons confined in the $(3,1)$ configuration at the Si--oxide interface in a large tunable QD and separated by a valley splitting, together with the relatively small D potential. The hyperfine interaction with the \ce{^31P} nucleus makes the electrons precess at different rates, creating an effective magnetic field difference $\Delta B_z = \pm A/2$ between the QD and the D.       \figletter{d} The donor electron can be moved to the QD using gate voltages. In this $(4,0)$ configuration, the exchange interaction dominates.     \figletter{e} Conceptual view of how a coupled QD and D cell could interact with other elements in a future chip. The electron qubit (blue arrows) is well suited for fast operations and readout. The nuclear spin qubit has high coherence and fidelity. Thanks to the large engineered QD, the electron qubits could be coupled through capacitive or exchange interaction without requiring atomic precision in the placement of donors.}
   \label{fig:fig1}
\end{figure*}

Here, we advance silicon-based quantum information processing by coherently coupling a phosphorus donor’s electron spin to a metal-oxide-semiconductor (MOS) QD. In our system, the QD is tuned to few-electron occupancy while simultaneously keeping a nearby donor (D) tunnel-coupled to the QD. The combination of the QD and donor electron qubits gives rise to a joint singlet-triplet (ST) logical encoding analogous to those in double-QD qubits \cite{levy2002,petta2005}. Specifically, the two logical states are the singlet $\ket{S}=\ket{\uparrow\downarrow}-\ket{\downarrow\uparrow}$ and unpolarized triplet $\ket{T_0}=\ket{\uparrow\downarrow}+\ket{\downarrow\uparrow}$. The encoding takes advantage of the contact hyperfine interaction between the donor electron spin and donor nuclear spin. This interaction makes the electron spin on the donor precess at a rate $A/2$ different from the QD electrons, where $A$ is the hyperfine coupling strength. The hyperfine interaction thus amounts to an effective magnetic field gradient produced by the single phosphorus nucleus and drives rotations between singlet and triplet states \cite{taylor2007}. By electrically controlling the donor charge configuration between ionized and neutral, the rotations can be turned off and on. The electron-electron exchange coupling and the hyperfine interaction with the donor nucleus define two orthogonal control axes for the qubit, and their relative strength is controlled using fast electrical pulses. 

The electron qubit formed by the QD-D coupled system is analogous to other ST qubits, while introducing important advantages. It features full electrical control with a uniquely compact design requiring only one QD. The QD-D ST qubit avoids the integration complexities of other Si spin control schemes such as micromagnets \cite{pioro-ladriere2008,wu2014a}, microwave striplines \cite{pla2012,veldhorst2014a} or additional QDs for full electrical control \cite{medford2013b,eng2015}. The hyperfine coupling to the single nuclear spin introduces a nature-defined and potentially very stable (i.e.\ low noise) rotation axis for the ST qubit. Furthermore, the system has a natural access to the nuclear spin, which is one of the highest performing solid state qubits \cite{muhonen2014b}. Integration of a coil for nuclear magnetic resonance could enable full control over the nuclear spin qubit. Nuclear spin readout schemes based on ST interactions with the donor have already been proposed \cite{kane2000}, making complete control of these two coupled qubits foreseeable in the near future.
The engineered coupling of the QD and D spins constitutes a possible path to realize over nineteen years of different theoretical proposals of donor qubit architectures \cite{kane1998,vrijen2000, skinner2003, calderon2009, tosi2017,pica2016}. For example, the large lithographic quantum dot can facilitate the coupling of neighboring QD-D cells using capacitive coupling \cite{levy2011,shulman2012a, nichol2017} or exchange interaction \cite{veldhorst2015a}.

\section*{Results}

\paragraph{Device description} 
The QD-D device is fabricated with isotopically-enriched \ce{^28Si} and a foundry-compatible process (i.e.\ no lift-off processing). We use a poly-silicon gate stack, shown in \figref{fig:fig1}a, that allows self-aligned ion implantation and subsequent activation annealing process. Phosphorus donors are implanted using the AG gate as a mask. This processing maximizes the probability of placing a D in a suitable location next to the QD. It also facilitates future multi-qubit fabrication that could take advantage of single ion implantation \cite{bielejec2010} and a planar QD geometry \cite{pica2016,tosi2017}. Fabrication details are found in the \RefSupSecFabAndMeasurement{} and are similar to Ref.\ \citenum{tracy2013}. A channel of electrons is formed at the MOS interface underneath the wire-shaped accumulation gate (AG) by applying a positive voltage, depicted as a blue overlay in \figref{fig:fig1}a. Next, a QD island is isolated by applying suitable negative voltages on neighboring gates. A single-electron transistor (SET) is formed in the upper wire to monitor the electron occupation $N$ of the QD and the relevant donor, denoted $(N_\text{QD},N_\text{D})$. The SET charge sensor (CS) is also used for spin readout via spin-to-charge conversion. An in-plane magnetic field of $300 \mT$ is applied throughout the experiments and the electron temperature is measured to be $215 \mK$. Detailed information about fabrication, gate biasing and electron counting is provided in the \RefSupSecFewElectrons{}.

To investigate coherent coupling dynamics between the donor and the QD, we first identify an effective $(2,0)\leftrightarrow(1,1)$ QD-D charge transition with a total of four electrons, as shown in \figref{fig:fig1}b-d \cite{nielsen2013, higginbotham2014b}. We use the spin filling structure, measured through magnetic fields, to engineer a sufficiently large energy difference $J_{(4,0)}$ between the singlet and triplet states \cite{johnson2005a}, which we observe to be substantially larger for four electrons ($\sim 150 \ueV$) than for two electrons ($\sim 60 \ueV$). Details are available in the \RefSupSecEffectiveTwoZero{}. In Si MOS, the valley splitting can be tuned to large values by increasing the electric field perpendicular to the interface, which was verified in this device \cite{gamble2016a}.  Simultaneously keeping the donor in resonance with the few electron QD states, however, constrained the available range of voltage in this design leading to the relatively small two-electron valley splitting.  We note two general benefits of using the four-electron configuration: (i) filled shells might be a general approach to circumvent the obstacle of low valley splitting in any material with conduction band degeneracy \cite{kawakami2014a,eng2015}; and (ii) increased electron numbers can extend the size of the QD due to the increased filling of the potential well, which in turn allows more range in selecting a suitable tunnel coupling to remote donor sites.

\begin{figure*}
   \centering
   \includegraphics{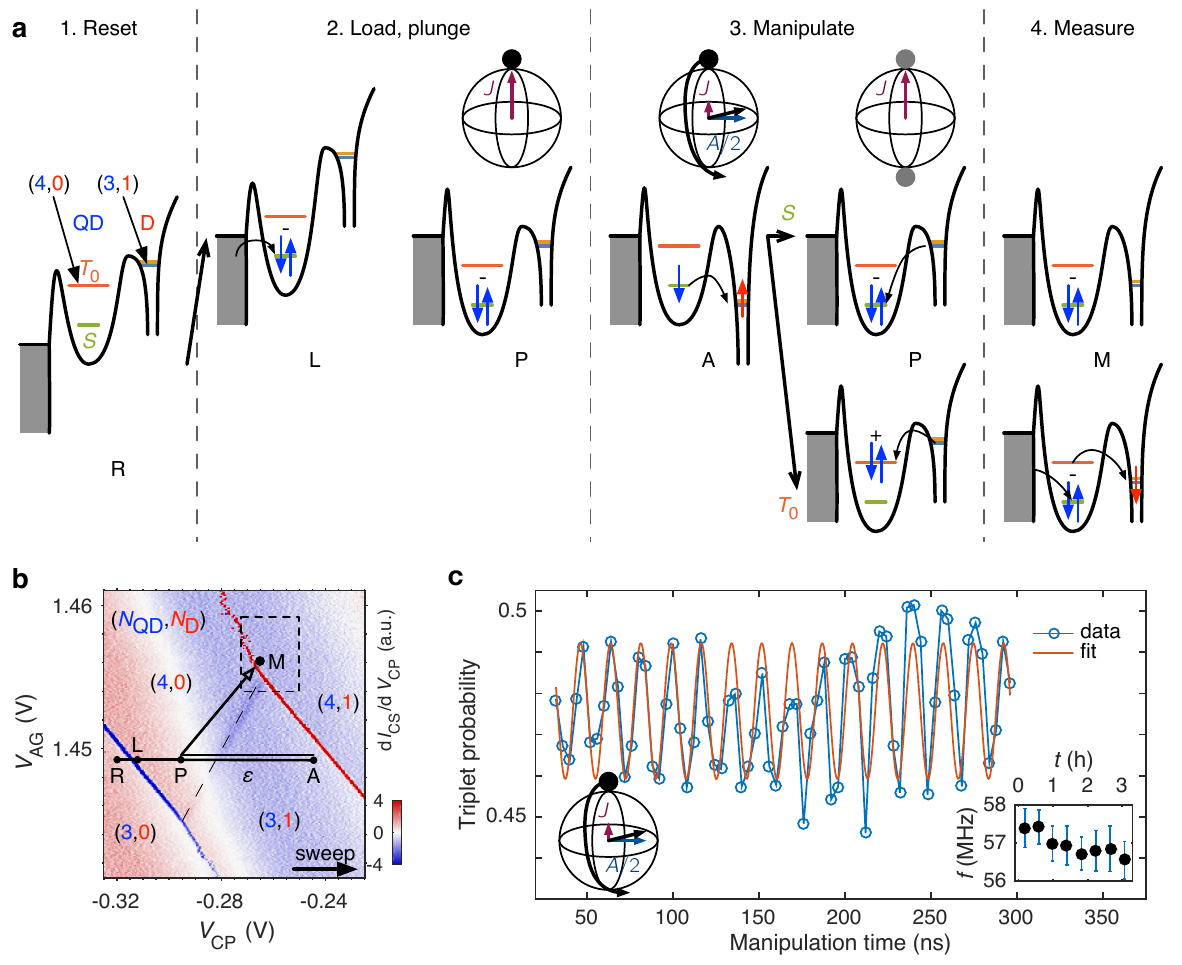} 
   \caption{ \figtitle{Hyperfine ST rotations.}      \figletter{a} Pulse sequence for the spin manipulations with schematic conduction band diagrams through the reservoir, QD and donor. The system is initialized as $(4,0)S$ and plunged to point P. Then, the detuning is pulsed rapidly to point A, which yields a separated $(3,1)S$. After a given manipulation time in $(3,1)$, which rotates the spin between $S$ and $T_0$, it is pulsed back to point P. The state is then either $(4,0)S$ or $(4,0)T_0$, and is measured by going to point M, where an enhancement in CS signal occurs (see \RefSupSecReadout{}).     \figletter{b} QD-D charge stability diagram. Overlaid are the different points of the experiment’s pulse sequence. The detuning $\epsilon$ is defined along the black line with zero detuning at the center of the $(4,0)\leftrightarrow(3,1)$ transition.     \figletter{c} Triplet probability versus manipulation time for $\epsilon = 950 \ueV$. The oscillation frequency is $f = 56.9 \MHz$. This is not the bare hyperfine frequency due to a residual exchange of $J/h = 27 \MHz$, see \figref{fig:fig3}c. Inset: Frequency of the oscillations for repeated measurements over 3 hours. Each point represents data averaged over 22 minutes, and the error bar represents the $95\pc$ confidence interval. }
   \label{fig:fig2}
\end{figure*}

\paragraph{Hyperfine-driven spin rotations}
Rotations between $\ket{S}$ and $\ket{T_0}$ can be driven by an effective magnetic field gradient $\Delta B_z = \pm A/2$ between the QD and the donor (in the remainder of the text we will drop the ket notation). These rotations provide a signature of the single \ce{^31P} donor. The source of the effective $\Delta B_z$ is the contact hyperfine interaction $A \scalar{\hat S}{\hat I}$ between the donor electron spin $\vct{\hat S}$ and the nuclear spin $\vct{\hat I}$. We expect the nuclear spin state to be projected onto a $\pm 1/2$ eigenstate by the repetitive experimental measurement. Rapidly separating a singlet state by pulling one electron onto the donor triggers coherent rotations between the $S$ and $T_0$ states. Reuniting the electrons onto the QD projects the state onto $S$ or $T_0$. We note that spin preparation, manipulations and readout act self-consistently with respect to a fixed but unknown state of the nuclear spin (i.e.\ the sign of $\Delta B_z$) in sufficiently large magnetic fields such that the interaction with the polarized triplets is suppressed (which is the case in this experiment). Moreover, nuclear states are known to be long lived ($\sim$ seconds) compared to the timescale of electron manipulations \cite{muhonen2014b}, therefore, errors caused by random flips while an electron is on the donor are expected to be negligible. The nuclear state could still have implications for single or multi-qubit operation. In the future, this could be addressed by deterministically setting the nuclear state through various pulsing schemes, such as a single-spin version of dynamic nuclear polarization \cite{nichol2015a}. To demonstrate the hyperfine-driven rotations, we use the pulse sequence shown in \figref{fig:fig2}. We prepare a $(4,0)S$ state by first emptying the QD and loading an electron between the singlet and triplet loading lines. Then, we plunge the system at point P (see \figref{fig:fig2}b). Next, we rapidly separate the electrons by pulsing the system to point A with a $16 \ns$ ramp time. After waiting for a given manipulation time, the system is pulsed back to point P in $(4,0)$. The ramp time is such that the charge transition is adiabatic, but fast enough to prepare a $(3,1)S$. Finally, we use an enhanced latching readout developed for this experiment and described in the \RefSupSecReadout{} to measure the triplet return probability. 

Figure \ref{fig:fig2}c shows the triplet return probability as a function of the manipulation time. All the details on the pulse sequence can be found in the \RefSupSecPulseAndCali{}. We find a ST rotation frequency $f = 57 \MHz$. This frequency is the vector sum of the exchange energy $J(\epsilon)$ and $A/2$, such that $hf = \sqrt{J^2+(A/2)^2}$, where $h$ is the Planck constant \cite{taylor2007}. We estimate a residual exchange of $J/h = 27 \MHz$ for this detuning from numeric fits to the frequency dependence on detuning (described below). The inset shows that this frequency is very stable over time. Such behavior differs from GaAs systems, for which dynamic nuclear polarization must be used to generate and maintain a particular $\Delta B_z$ of similar magnitude \cite{barthel2012}. The magnitude and stability of $f$ provides a strong indication that the rotations are driven by a single \ce{^31P}. A small and relatively constant frequency drift of around $0.8 \MHz$ is observed over a period of $3.5 \text{ h}$ which is consistent with the drift in the electrostatics of the device through the experimentally measured $\D J/\D \epsilon$ relation. Additionally, the observed linewidth is less than natural silicon, which has linewidths greater than $8 \MHz$ for single donors \cite{pla2012} and is qualitatively consistent with an enriched \ce{^28Si} background. Noise in $J$ is believed to presently limit the linewidth, discussed below in terms of $T_2^*$.

\begin{figure}
   \centering
   \includegraphics{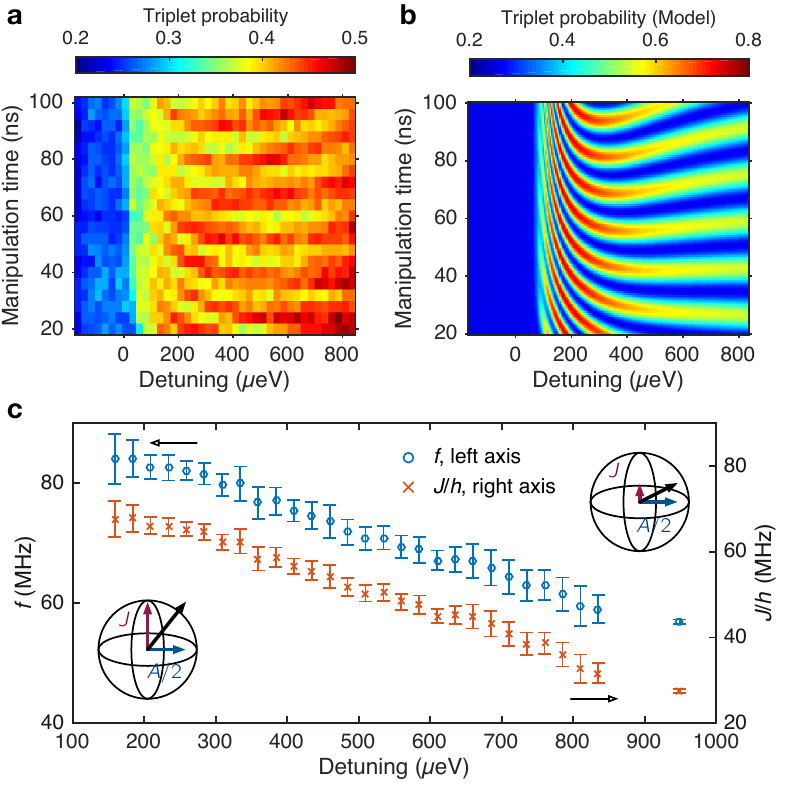} 
   \caption{ \figtitle{Detuning dependence of exchange.}     \figletter{a-b} Experimental (a) and model (b) triplet probability versus detuning and manipulation time. The oscillations on the time axis are the hyperfine-driven rotations. The phase and shape of the oscillations are well reproduced; however, the mechanisms limiting the visibility are detailed in the \RefSupSecVisibility{}.     \figletter{c} Left axis: Frequency $f$ of the oscillations versus detuning, extracted from fits to \figref{fig:fig3}a. This corresponds to the $S$--$T_0$ energy gap, allowing a fit for model parameters $t_\text{S} = 19 \ueV$, $t_\text{T} = 31 \ueV$ and $A/2h = 50 \MHz$. Error bars represent the $95\pc$ confidence interval. Right axis: Exchange $J$ calculated from the frequency after removing the hyperfine contribution $J/h = \sqrt{f^2-(A/2h)^2}$. }
   \label{fig:fig3}
\end{figure}

\paragraph{Characterization of exchange interaction} 
The detuning dependence of the ST rotations reveals additional information about this QD-D system. In \figref{fig:fig3}a, we plot the triplet return probability against both detuning and manipulation time. As the detuning gets closer to zero, the frequency of the exchange rotations increases, as shown in \figref{fig:fig3}c. This is consistent with a ST model where the exchange energy $J$ between the $S$ and $T_0$ states is not negligible and drives rotations around a tilted axis in the qubit Bloch sphere. To better understand the exact shape of the oscillations of \figref{fig:fig3}a, we simulate the quantum dynamics of the system using a master equation approach and time-dependent controls. We describe the system using the basis states $\{(4,0)S, (4,0)T_0, (3,1)S, (3,1)T_0\}$, similarly to previous treatments such as Taylor \etal{}\ \cite{taylor2007}. The details of the model are given in the \RefSupSecModel{}. The numerical simulation results are shown in \figref{fig:fig3}b. The phase and shape of the oscillations is very well reproduced; however, the mechanisms limiting the visibility are numerous and detailed in the \RefSupSecVisibility{}. At the moment, we think that addressing the various causes could ultimately produce results on-a-par or even better than state-of-the-art ST qubits. The key fitting parameters of the model are the triplet tunnel coupling $t_\text{T}$, singlet tunnel coupling $t_\text{S}$, and hyperfine interaction $A$. We can determine these parameters using a fit to the data of \figref{fig:fig3}c, knowing that $hf$ equals the energy gap between the $S$ and $T_0$ states. We find $t_\text{S} = 19 \ueV$, $t_\text{T} = 31 \ueV$ and $A/2h = 50 \MHz$. Shifts in $A$ of this magnitude relative to the bulk value (of $58.5 \MHz$, Ref.\ \citenum{pla2012}) have been reported in single donor electron spin resonance (ESR) experiments \cite{muhonen2014b} and have been attributed to Stark shifts of the contact hyperfine interaction due to the large electric fields in the vicinity of the neighboring QD. The measured value in this work is both consistent with a shallow phosphorus donor and is inconsistent with likely alternatives, such as arsenic. Following the fit procedure, we can extract $J(\epsilon)$ by subtracting the $A/2$ contribution. The result is shown in \figref{fig:fig3}c.

\section*{Discussion}

\begin{figure}
   \centering
   \includegraphics{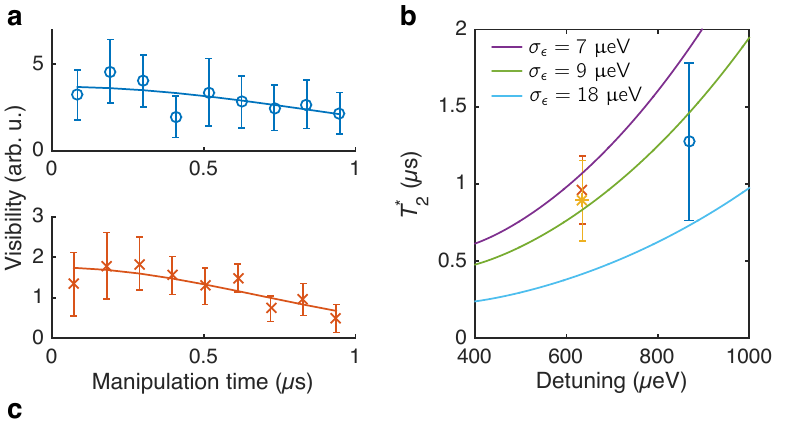} \\
   \small
   \begin{tabular}{c|cccc}
   	\hline\hline
   	Reference	& Petersson \etal	& Shi \etal	& Eng \etal	& This work \\
	\hline
	Material 	& GaAs/AlGaAs 	& Si/SiGe 	& Si/SiGe 	& Si (MOS) \\
	$2\sigma_\epsilon$ ($\upmu$eV)	& 7.4	&10	& 9.2	& 18 \\
	$k_\text{B}T_\text{e}$ ($\upmu$eV)	& 6.9	& 12	& 6.9 -- 8.6	& 18 \\
	\hline\hline
   \end{tabular}
   \caption{ \figtitle{Coherence time.}     \figletter{a} Visibility of ST rotations versus time and $T_2^*$ fit with a Gaussian decay model for $\epsilon = 868 \ueV$ (top) and $635 \ueV$ (bottom). The results are plotted in (b). Error bars represent the $95\pc$ confidence interval.     \figletter{b} The $T_2^*$ at both detunings is consistent with residual exchange and a detuning noise of $\sigma_\epsilon = 9 \ueV$. Error bars represent the $95\pc$ confidence interval.     \figletter{c} Table comparing charge noise in a variety of material systems. All show a similar noise level that seems related to the electronic temperature $T_\text{e}$. $k_\text{B}$ is the Boltzmann constant. References: Petersson \etal\ \cite{petersson2010}, Shi \etal\ \cite{shi2013}, Eng \etal\ \cite{eng2015}. }
   \label{fig:fig4}
\end{figure}

Decoherence of MOS QDs \cite{veldhorst2014a} and single donors \cite{muhonen2014b} has been characterized in separate systems, but the charge noise and magnetic noise properties of strongly hybridized QD-D systems are not well established. Our system provides a unique platform to study these important properties in an effective two-electron case where entanglement is delocalized in the form of a spatially separated singlet or triplet. We measure long time traces and plot the visibility of the oscillations versus manipulation time $t$ in \figref{fig:fig4}a. The data and method are presented in the \RefSupSecCoherenceTime{}. We then fit the decay using a slow detuning noise model that produces a Gaussian decay of the visibility $v = v_0 \Exp{-(t/T_2^*)^2}$, where $v_0$ is an arbitrary initial visibility. We find that $T_2^*$ depends on the detuning (\figref{fig:fig4}b). To understand this dependence, we use a charge noise model represented by $\epsilon$ noise with a characteristic standard deviation $\sigma_\epsilon$ and producing decoherence through $J(\epsilon)$ \cite{dial2013}. Details about the model are given in the \RefSupSecNoise{}. We find that $\sigma_\epsilon = 9 \ueV$ is consistent with the observed $T_2^*$. In this model, we neglect magnetic noise that could be caused by residual \ce{^29Si} or other sources. Our observations are consistent with $T_2^*$ being limited by charge noise, a mechanism that is expected to play an important role when $J$ varies as a function of $\epsilon$ \cite{dial2013}. We note that $2\sigma_\epsilon$ is approximately the electronic temperature $k_\text{B}T_\text{e}$. The noise magnitude has previously been correlated with the electronic temperature \cite{dial2013}. We further tabulate noise magnitudes in a variety of material systems, like GaAs/AlGaAs heterostructures \cite{petersson2010}, Si/SiGe heterostructures \cite{shi2013,eng2015} and MOS (this work), and show the results in \figref{fig:fig4}c. 

In summary, we have demonstrated coherent coupling between the electrons of two very different qubit systems: a donor atom (natural atom) and a MOS quantum dot (artificial atom \cite{kastner1993}). The coherent rotations between the singlet and triplet are driven by a nuclear spin qubit through the contact hyperfine interaction, and produce $10 \ns$ $X(\pi)$ rotations with a $T_2^*$ of $1.3 \pm 0.7 \us$, thus allowing over 100 rotations within the coherence time. A charge noise magnitude of $9 \ueV$ fits the stationary noise model and is a characterization of the MOS interface noise properties, which are found to be of similar magnitude to other common QD material systems. Assuming this model, the $T_2^*$ could possibly be improved by a factor 10 or more by operating at larger detunings where the exchange is negligible, hence taking full advantage of isotopically pure silicon. Our experiments demonstrate the feasibility of using the QD-D system as a compact ST qubit with no additional micromagnets \cite{pioro-ladriere2008,wu2014a} or QDs (as in all-exchange qubits \cite{gaudreau2012,medford2013a,eng2015}), and avoid the decoherence mechanisms associated with GaAs or Si host nuclear species \cite{petta2005,maune2012}. More sophisticated ST qubit control approaches \cite{shulman2014,cerfontaine2014} and optimized preparation/readout parameters will likely increase the visibility and reduce errors of future two-axis QD-D qubit demonstrations. To further speed up the operations compared to the coherence time, it could be possible to use other donor species that have stronger contact hyperfine strengths. Beyond individual ST qubits, this work opens-up compelling possibilities. One such example is the coupling of donor-based qubits without atomic precision placement through, for example, electrostatic coupling between ST qubits \cite{levy2011,shulman2012a,nichol2017}. Another example is all-electrical nuclear spin readout \cite{kane2000} and electric/nuclear magnetic resonance control without high magnetic fields or ESR, thus introducing a nuclear spin qubit as an additional resource.

\bibliographystyle{naturemag} 
\bibliography{/Users/Patrick/Documents/Papers/PHC} 

\section*{Acknowledgements}

The authors would like to thank Erik Nielsen, Andrew D.\ Baczewski, Matthew J.\ Curry and Stephen Carr for valuable help and discussions regarding this work; B.\ Silva for support on device fabrication and room temperature measurements; and Benjamin D’Anjou and William A.\ Coish for discussions on readout fidelity. This work was performed, in part, at the Center for Integrated Nanotechnologies, an Office of Science User Facility operated for the U.S.\ Department of Energy (DOE) Office of Science. Sandia National Laboratories is a multimission laboratory managed and operated by National Technology and Engineering Solutions of Sandia, LLC, a wholly owned subsidiary of Honeywell International, Inc., for the U.S.\ DOE's National Nuclear Security Administration under contract {DE-NA0003525}.

\section*{Author Contributions}

P.H.-C.\ and M.S.C.\ designed the experiments. 
P.H.-C.\ performed the central measurements and analysis presented in this work. 
M.R.\ performed supporting measurements on similar “control” samples that establish repeatability of many observations in this work. 
N.T.J., P.H.-C., M.R.\ and J.K.G.\ modelled key elements of the device structure providing critical insights. 
P.H.-C., M.S.C., N.T.J.\ and M.P.-L.\ analyzed and discussed central results throughout the project, including designing models for observations. 
J.D., T.P., G.A.T.E.\ and M.S.C.\ designed process flow, fabricated devices and designed/characterized the \ce{^28Si} material growth for this work. 
J.R.W.\ provided critical nanolithography steps. 
M.L.\ supplied critical laboratory set-up for the work. 
M.S.C.\ supervised combined effort including coordinating fab and identifying modelling needs for experimental path. 
P.H.-C., M.S.C.\ and M.P.-L.\ wrote the manuscript with input from all co-authors.

\section*{Additional information}
\noindent
\textbf{Supplementary Information} is included with this paper. 
\\ \\ \noindent
\textbf{Competing financial interests:} The authors declare no competing financial interests. 
\\ \\ \noindent
\textbf{Reprints and permission} information is available online at http://npg.nature.com/reprintsandpermissions/
\\ \\ \noindent
\textbf{Data availability:} The authors declare that the data supporting the findings of this study are available within the paper and its supplementary information files. Additional data (e.g.\ source data for figures) are available from the corresponding author upon reasonable request.



\makeatletter
\renewcommand\thesection{\mbox{S\arabic{section}}}
\makeatother
\setcounter{section}{0}     

\newcounter{supfigure} \setcounter{supfigure}{0} 
\makeatletter
\renewcommand\thefigure{\mbox{S\arabic{supfigure}}}
\makeatother

\newcounter{suptable} \setcounter{suptable}{0} 
\makeatletter
\renewcommand\thetable{\mbox{S\arabic{suptable}}}
\makeatother

\newenvironment{supfigure}[1][]{\begin{figure}[#1]\addtocounter{supfigure}{1}}{\end{figure}}
\newenvironment{supfigure*}[1][]{\begin{figure*}[#1]\addtocounter{supfigure}{1}}{\end{figure*}}
\newenvironment{suptable}[1][]{\begin{table}[#1]\addtocounter{suptable}{1}}{\end{table}}
\newenvironment{suptable*}[1][]{\begin{table*}[#1]\addtocounter{suptable}{1}}{\end{table*}}

\newcommand{\RefFigOne}{\mbox{\figref{fig:fig1}}} 
\newcommand{\RefFigTwo}{\mbox{\figref{fig:fig2}}} 
\newcommand{\RefFigThree}{\mbox{\figref{fig:fig3}}} 
\newcommand{\RefFigFour}{\mbox{\figref{fig:fig4}}} 

\clearpage
\newpage
\onecolumngrid
{\centering
\large\textbf
{Supplementary Information for: \\ \mytitle} \\ \rule{0pt}{12pt}
}
\twocolumngrid

\section{Device Fabrication and Measurement} \label{sec:fabmeas}

The device used for these experiments is fabricated identically to the one of \RefFigOne{}a. Electrons are confined in a 2D electron gas at the interface between an epitaxial enriched \ce{^28Si} layer with $500 \text{ ppm}$ residual \ce{^{29}Si} and a $35 \nm$ gate oxide. Highly n-doped poly-silicon gates ($200 \nm$ thick) are patterned on top of the gate oxide using low pressure chemical vapor deposition and plasma etching \cite{tracy2013}. These are used to accumulate electrons by applying a positive voltage (in an enhancement mode) or deplete electrons (with negative voltages). Phosphorus donors are implanted in a PMMA resist window that overlaps with the AG gate on both sides of both wires, and the poly-Si gate used as a self-aligned implantation mask. The approximate relevant donor location indicated by the red dot in \RefFigOne{}a of the main text is inferred from various donor-gate capacitance ratios. The source and drain reservoir electrons are connected by n+ regions and ohmic contacts to the instruments. The device is biased to form a SET in the upper wire that is used as a charge sensor (CS), while simultaneously forming a few-electron QD under the lower wire. The CS current $I_\text{CS}$ is measured using an AC lock-in technique at $403 \Hz$ with $0$ DC source-drain bias and $100 \uV$ (rms) AC bias. The derivative with respect to gate voltage is taken numerically to show the QD charge occupancy steps in charge stability diagrams.

\section{Few electron regime} \label{sec:fewel}

One can form a clean single QD with this device geometry through biasing that pushes the QD towards one lead, shown in \figref{fig:figs1}a. A representative set of gate voltages used for this experiment is shown in \figref{fig:figs1}c. We use gates AG and CP to discriminate between QD and D states, respectively. The region where donors interact resonantly with the QD is shown in \figref{fig:figs1}b. In this regime, the single QD turns into two strongly coupled QDs in series along the wire axis. This is indicated in the charge stability diagram of \figref{fig:figs1}b by two sets of nearly parallel lines. This behavior is systematically reproduced in the devices we measured with such a geometry, which indicates that it is a feature produced by the electrostatics of the device. We can assign occupation numbers to the two QDs, counting from zero. We establish that the QD is in the few electron regime (i.e.\ emptied) by opening the tunnel barriers to the point where the QD charge-sensed lines become lifetime-broadened without detecting other states (data not shown). Donor and/or defect transitions can be seen cutting through the QD lines and are identified by red lines. These objects anti-cross with the QD lines in a way analogue to double QDs. The main difference is that they can only accommodate a limited number of charge states, like 0 or 1. We additionally performed magnetospectroscopy \cite{tarucha1996,potok2003,xiao2010a} to verify that the first electron fills as a spin-down electron. For this work, we treat the QD closest to the reservoir as being part of the reservoir itself and neglect its impact on the other QD.
\begin{supfigure*} 
   \centering
   \includegraphics{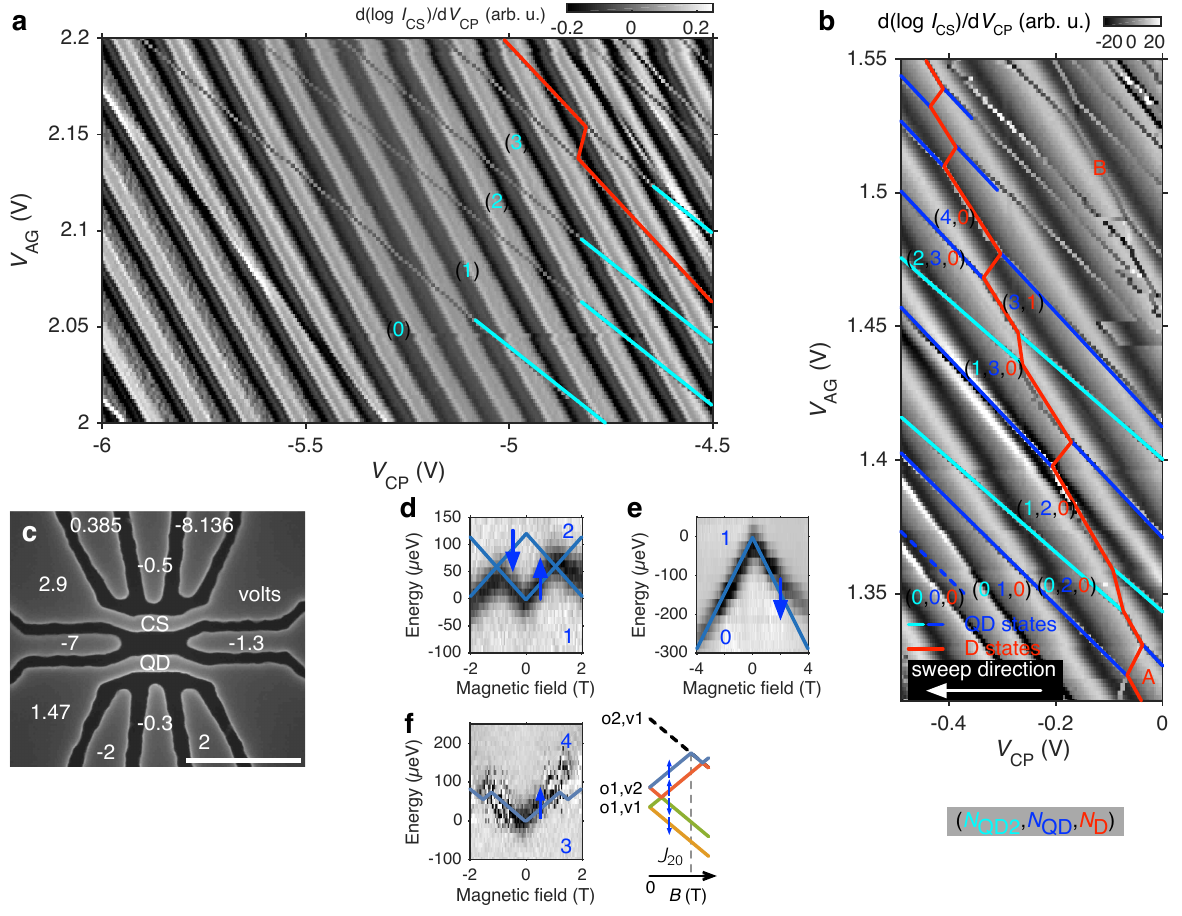} 
   \caption{ \figtitle{Quantum dot and donor states.}    \figletter{a} Charge stability diagram in the single dot regime. The broad background features are the successive Coulomb peaks of the SET charge sensor (CS). The sharp features indicated by colored lines are the QD and donor charge transitions.      \figletter{b} Charge stability diagram showing all states. Notation: $(N_\text{QD2},N_\text{QD},N_\text{D})$. The QD2 represents a QD that is strongly coupled to the lead and is ignored in the main text. The red lines represent donor or defect states cutting through the QD states. These anti-cross with the QD in a double-QD honeycomb fashion but have only 0 or 1 electron occupancies possible.     \figletter{c} Typical gate voltages for the experiment.      \figletter{d-e} Opening the QD-lead tunnel barriers allows to probe the spin filling of the first two QD states through magnetospectroscopy. CP gate voltage has been converted to energy using a lever arm.     \figletter{f} Magnetospectroscopy data showing the $(3,0)\leftrightarrow(4,0)$ transition loading as a spin singlet. The $(4,0)$ ground state hence forms an effective $(2,0)S$ with exchange splitting $J_{(4,0)} = 143 \ueV$ (confirmed with pulse spectroscopy). The CP gate voltage was converted to energy $E$ through a lever arm. Grey scale: $\D I_\text{CS}/\D E$ (arb.\ u.). Right schematic: The observed spin filling is qualitatively consistent with a simple shell filling model (see e.g.\ Ref.\ \citenum{lim2011}). The states have a valley-like (v1 and v2) or orbit-like (o1 and o2) character.}
   \label{fig:figs1}
\end{supfigure*}

\section{Effective (2,0)-(1,1) system} \label{sec:effectivedouble}

To investigate singlet-triplet dynamics, we first identify an effective $(2,0) \leftrightarrow (1,1)$ QD-D charge transition with a total of four electrons, as shown in \RefFigTwo{}b. Singlet-triplet states with more than two electrons have been studied theoretically \cite{nielsen2013} and experimentally \cite{higginbotham2014b} in double-QD systems. Using magnetospectroscopy \cite{tarucha1996,potok2003,xiao2010a}, we verify that the QD spin filling is indeed consistent with having a four-electron singlet ground state (see \figref{fig:figs1}f). A requirement for efficient spin initialization and readout is that the energy difference $J_{(2,0)}$ between the singlet $(2,0)S$ and triplet $(2,0)T_0$ be much larger than the electron temperature of the experiment \cite{johnson2005a}, which is $215 \mK$ in this case. Hence charge transitions have a full width at half maximum of approximately $65 \ueV$. In silicon, the valley splitting is generally the factor limiting $J_{(2,0)}$ \cite{lai2011}. In our device and for the values of $V_\text{AG}$ used, we have measured the valley splitting to be approximately $60 \ueV$. Consistent with this observation, the two-electron QD states had similarly small values for $J_{(2,0)}$. The four-electron QD state of \RefFigTwo{}b, however, has an appreciably larger ST splitting of $J_{(4,0)} = 143 \ueV$ (as measured from both magnetospectroscopy and excited state spectroscopy). This might be understood as a shell filling effect with QD orbitals, where the pairing of spins allows to circumvent the small valley splitting, as illustrated in the schematic of \RefFigOne{}c-d \cite{lim2011}.

\section{Spin preparation and readout} \label{sec:spinprepread}

We show that we can initialize and read out ST spin states. To do so, we use the pulse sequence of \figref{fig:figS2}a. The system is initialized into a $(4,0)S$ or $(4,0)T$ state (where $T$ stands for any triplet) by first ejecting the fourth electron at point R (as defined in \RefFigTwo{}b), and then loading either a singlet (S) or triplet (T) state by carefully tuning the load level of point L. A deeper load tends to prepare T states due to their $\sim 10$ times faster loading rate. After passing through an intermediate point P, which will be important for spin manipulations, the gate voltages are pulsed to point M for spin readout. The readout mechanism is shown in \figref{fig:figS2}b. Through Pauli-blockade, the spin state is converted to either a $(4,0)$ or $(4,1)$ charge state depending on whether the initial spin state was a singlet or a triplet, respectively. The mechanism relies on a charge hysteresis effect caused by the absence of direct access to a charge reservoir for the donor (\figref{fig:figS2}c). Hence, the donor\ $\leftrightarrow$\ lead transitions are very slow because they have to go through a co-tunneling process to equilibrate \cite{yang2014a}. Placing point M between the S and T charge preserving transitions then allows a fast relaxation path to the charge ground state only if the initial state was $(4,0)T$. If the state was $(4,0)S$, the system is locked in a metastable charge configuration. The resulting CS signal is enhanced because the final charge configuration differs by one electron and lasts longer than the relaxation time of the $(4,0)T$ state. A charge enhancement effect like this has been previously highlighted by Studenikin \etal{}\  \cite{petersson2010,studenikin2012a}. This readout mechanism allows us to use averaged measurements instead of single-shot. Since the measurement step is the longest in the pulse sequence, the current at point M in \figref{fig:figS2}d-e is proportional to the triplet probability. All state measurements throughout this work are averaged over many (150 to 200) cycles. Details about the pulse sequence, loading rates, relaxation rates and probability calibration are given in the next section. 
\begin{supfigure*}
   \centering
   \includegraphics{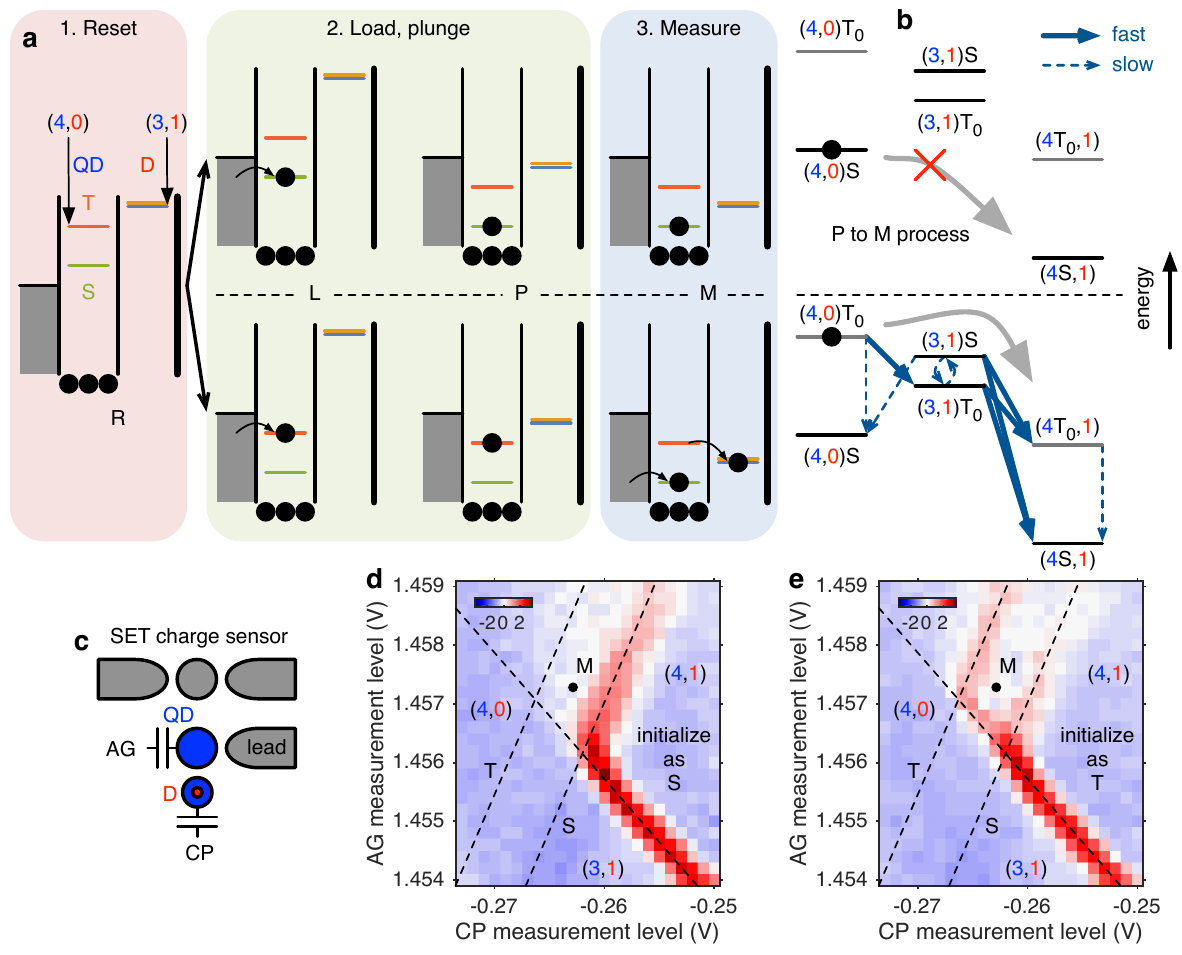} 
   \caption{ \figtitle{Spin preparation and readout.}     \figletter{a} Measurement pulse sequence. The state is reset by emptying the QD at point R (see \RefFigTwo{}b). A $(4,0)S$ or $(4,0)T$ is loaded at point L. After passing through point P, the system is biased to point M for readout.     \figletter{b} Readout process at point M. If the state was $(4,0)S$, the system is locked in a metastable charge state. If the state was $(4,0)T$, the pulse is beyond the $(4,0) \leftrightarrow (3,1)$ transition line, which unlocks the relaxation to $(4,1)$ by going through $(3,1)$.       \figletter{c}~Schematic showing the QD-D configuration. The D has no direct connection to a lead, which causes the hysteresis.     \figletter{d-e}~Readout demonstration. The coordinates of point M (only) are swept across the charge stability diagram. When preparing predominantly singlet (S) states, as in (D), only the S QD-D charge transition is visible. In contrast, preparing predominantly triplet (T) states reveals the T QD-D charge transition, as in \textbf{e}. The CS current at the location labeled M is then proportional to the triplet probability. Color scale: $\D I_\text{CS}/\D V_\text{CP} \text{ (arb. u.)}$. }
   \label{fig:figS2}
\end{supfigure*}

\section{Pulse sequence, loading and relaxation rates} \label{sec:pulseseq}

The AC component of pulses in the experiment is applied using an Agilent 33500B arbitrary waveform generator using two synchronized channels for the AG and CP gates. The waveform is composed of DC and AC components and applied to the gates through a room temperature bias tee. The waveforms are applied such that all target points are fixed in the charge stability diagram, except the ones explicitly varied for a particular measurement (e.g.\ manipulation time or position of point M). The $(4,0)S$ loading rate is approximately $1/(60 \us)$, and the $(4,0)T$ loading rate approximately $1/(6 \us)$. The $(4,0)T-(4,0)S$ relaxation time is approximately $375 \us$, determined by preparing mostly $(4,0)T$ and measuring the triplet probability decay versus time. The metastable state lifetime is roughly 2 to $4 \ms$. We define zero detuning (the energy difference $\epsilon$ between the QD and D) at the QD-D charge transition, and positive detunings along $V_\text{CP}$ in the $(3,1)$ direction with a $17 \ueV\text{\,mV}^{-1}$ lever arm. Then, we plunge the system to $\epsilon = -250 \ueV$ at point P (see \RefFigTwo{}b). Next, we rapidly pulse the system to $\epsilon = 950 \ueV$ (point A, \RefFigTwo{}) or a variable detuning (\RefFigThree{}) with a $16 \ns$ ramp time. After waiting for a given manipulation time, the system is pulsed back to point P in $(4,0)$.
\begin{suptable}
\caption{ \figtitle{Pulse sequence parameters.}     Table of pulse sequence points (as defined in main text \RefFigTwo{}b), ramp time to point (from previous point), and wait time at point, for a typical manipulation pulse sequence used. The sequence is played in a loop.}
\begin{center}
\begin{tabular}{c|r|r}
	\hline\hline
	Point	& Ramp time ($\upmu$s)	& Wait time ($\upmu$s) \\
	\hline
	R		& 10					& 50 \\
	L		& 0.1				& 150 \\
	P		& 1					& 0.2 \\
	A		& 0.016				& 0.1 \\
	P		& 0.016				& 0.2 \\
	M		& 10					& 350 \\
	\hline\hline
\end{tabular}
\end{center}
\label{tab:tabs1}
\end{suptable}

\section{Probability calibration} \label{sec:probcal}

To calibrate the triplet probability, the following procedure is used. First, the CP gate voltage of the measurement point M and the loading point L are swept to tune the readout and initialization, respectively, using the same waveform as for state manipulation except for point A (such that no manipulations are done). The resulting CS current is mapped in \figref{fig:figs3}a. Given a certain load level, the CS current is then plotted versus CP measurement level, \figref{fig:figs3}b. The current has a downward linear trend because of the CS Coulomb peak flank and a step that is similar in origin to a normal charge sensing signal. To the left of the measurement window the current always corresponds to a singlet signal, and to the right it always corresponds to a triplet signal. By extrapolating what this current would be assuming a linear background, one can determine what the pure singlet and triplet signals should be in the measurement window. The actual triplet probability is determined using a linear transformation that maps $I_\text{CS}$ to triplet probability. When manipulations are performed, the duty cycle of the waveform is changed by at most $0.2\%$, so the calibration is largely unaffected. Any systematic error introduced by this method (e.g.\ broadening of transitions due to temperature) would tend to underestimate the visibility of oscillations.
\begin{supfigure} 
   \centering
   \includegraphics{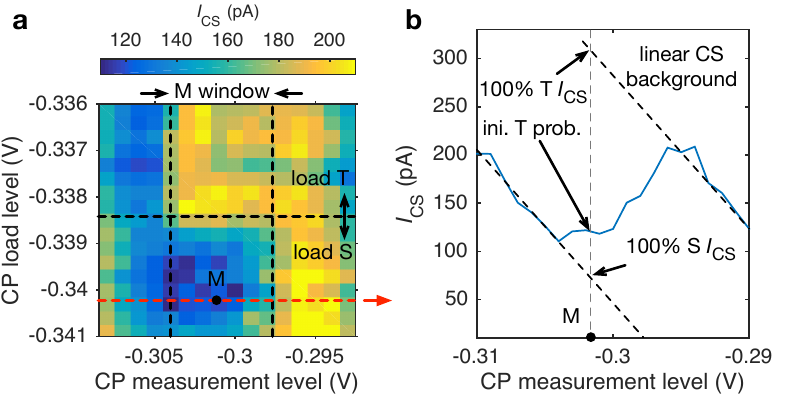} 
   \caption{ \figtitle{Probability calibration.}     \figletter{a} The CP gate voltage of measurement point M and loading point L are swept to tune the readout and initialization, respectively, using the same waveform as for state manipulation except for point A (no manipulations). This method allows to map the measurement and initialization windows.     \figletter{b} A cut through measurement levels reveals what the current would be for pure singlet (S) or triplet (T) states in the measurement window.}
   \label{fig:figs3}
\end{supfigure}

\section{Singlet-triplet dynamics model} \label{sec:stmodel}

We model the ST system with a $4\times4$ Hamiltonian in the basis $\{ \ket{(4,0)S}, \ket{(4,0)T_0}, \ket{(3,1)S}, \ket{(3,1)T_0} \}$ given by
\ma{
	H(t) =
	\frac12 
	\mat{\eps(t) & 0 & -t_\text{S} & 0 \\
	0 & 2J_{(4,0)}+\eps(t) & 0 & -t_\text{T} \\
	-t_\text{S} & 0 & -\eps(t) & -A/2 \\
	0 & -t_\text{T} & -A/2 & 2J_{(3,1)}-\eps(t) }  ,
}
where $J_{(4,0)}$ and $J_{(3,1)}$ are the exchange between singlet and triplet states in the $(4,0)$ and $(3,1)$ charge sectors, respectively, $t_\text{S}$ ($t_\text{T}$) is the QD-D tunnel coupling for the singlet (triplet) states, $A/2$ is the effective magnetic field gradient due to the contact hyperfine interaction in the $(3,1)$ configuration, and $\eps(t)$ is the detuning. We fix $J_{(4,0)} = 143 \ueV$ and $J_{(3,1)} = 0 \ueV$. For a given control schedule $\eps(t)$, we numerically integrate to solve for the time evolution of the density matrix $\rho(t)$ generated by the von Neumann equation
\ma{
	\deriv{\rho(t)}{t} = -\frac{i}{\hbar} \comm{H(t)}{\rho}   .
}
To model the effect of finite control bandwidth, the pulse sequence we consider in our numerical simulations is given by the ideal pulse sequence after having been filtered through a (low pass) RC filter, 
\ma{
	\tilde\eps(t) = \int_{-\infty}^{\infty} \D\tau h_\text{RC}(\tau) \eps(t-\tau)   ,
}
where $h_\text{RC}(\tau) = \frac{1}{RC} \theta(t) \E{-t/RC}$ is the impulse response function and $\theta(t)$ is the Heaviside step function. Considering various filtered control schedules $\tilde\eps(t)$, we find that a time constant $RC = 10 \ns$ is consistent with the experiment.

\section{Factors limiting the visibility} \label{sec:factorsvisi}

In the main text \RefFigTwo{}c, the visibility of the coherent rotations is low. This discussion identifies the different contributions to the visibility. It should be noted that the rotations are approximately 100 times faster than the coherence time. Therefore, the fidelity of the rotation itself should be quite high. Factors contributing to the reduced visibility are state preparation and measurement (SPAM) errors, additional incoherent or leakage processes during the fast ramp in/out of the $(3,1)$ region, and the control protocol itself. It should be noted that the control protocol used is not expected to produce full visibility according to our simulations. This is in part due to the limited bandwidth of the pulse in this setup (i.e.\ part of the wavefunction remains in the ground state because of partial adiabatic transfer of the spin state in the strong gradient field). Through various measurements we estimate that preparation errors alone are responsible for the $24\pc$ triplet probability background in the $(4,0)$ region and limit the visibility to $(1-0.24\times2) = 52\pc$. Singlet preparation was limited by the slow QD-lead tunnel rate which required long loading steps that were competing with the bias tee time constant. The readout process could also yield additional errors at the $\sim15\pc$ and $\sim~30\pc$ levels for singlets and triplets respectively due to triplet relaxation and various technical compromises. The dynamics model in the main text takes preparation errors into account and predicts a visibility of approximately $30\pc$, which is the simulated data shown in the main text. Adding measurement errors further reduces the expected visibility to $\sim 17\pc$. In the main text data, the visibility is around $6\pc$. This additional loss of visibility is dominated by an error process that occurs when the zero-detuning line is crossed. The exact mechanism is unknown. We speculate that it could be due to incoherent charge excitation/relaxation near the zero detuning point itself. If this is the case, a faster pulsing rise time and a bigger tunnel coupling would be expected to reduce errors.

\section{Coherence time analysis} \label{sec:T2analysis}

\begin{supfigure*}
   \centering
   \includegraphics{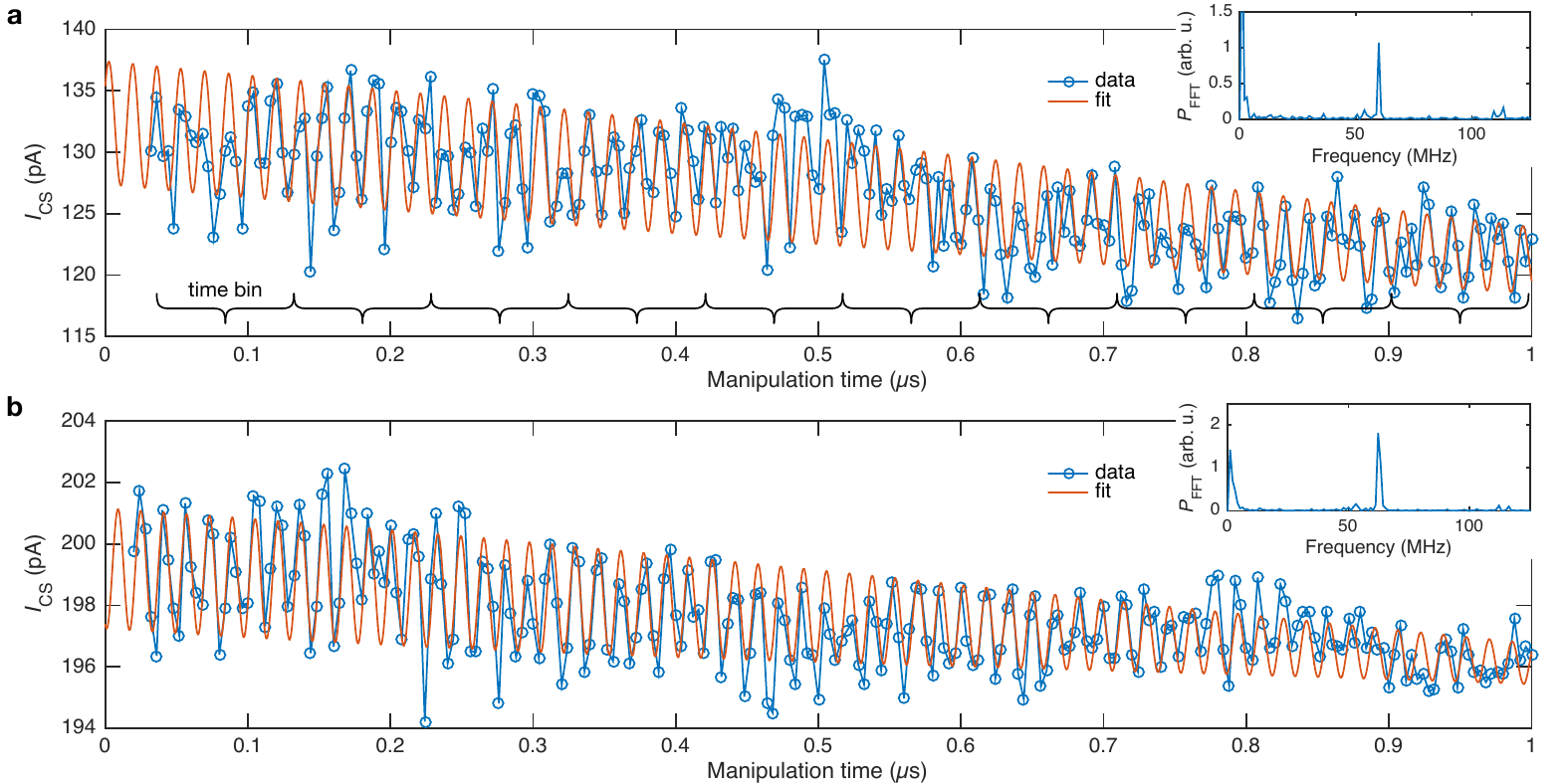} 
   \caption{ \figtitle{Long singlet-triplet rotations.}     \figletter{a-b} Singlet-triplet oscillations used to extract the visibility and $T_2^*$. The charge sensor current $I_\text{CS}$ is proportional to the triplet probability. The visibility as a function of time is determined by the amplitude of a sine fit for each time bin separately (not shown). The overlay fit curve is a sinusoidal fit with Gaussian decay that has good agreement with the data and the time binning method. In (a), $\epsilon = 868 \ueV$ and the fit yields $T_2^* = 1.3 \pm 0.7 \us$. The error is the $95\pc$ confidence interval. In (b), $\epsilon = 635 \ueV$ and the fit yields $T_2^* = 0.96 \pm 0.31 \us$. Insets: Fast Fourier Transform power ($P_\text{FFT}$) spectrum showing a clear single-frequency signal.}
   \label{fig:figs4}
\end{supfigure*}

To extract the visibility $v$ of the ST oscillations of \RefFigFour{}a, the following method is employed. The source data is shown in \figref{fig:figs4}. Because of CS drift over the long periods of time required to acquire these longer time traces (2 hours each in this case), the CS current (proportional to triplet probability) has a general downward trend and some residual fluctuations. To remove these fluctuations and smooth the data, the time trace is divided into time bins of approximately $100 \ns$. The oscillations in each time bin are fitted with a sine function of fixed frequency. The amplitude for each time bin is then reported as visibility in \RefFigFour{}a. The visibility decay is then fitted using a Gaussian decay, as detailed in the “Detuning noise model” section. We have verified that this time binning method agrees well with other methods such as maximum likelihood analysis. We now look at the apparent modulations of the oscillations in \RefFigTwo{}c of the main text and \figref{fig:figs4}. These are believed to arise from the averaging of a limited ensemble of traces with slightly different frequencies. This is expected because of the slow charge noise and light drift, and leads to beating-like features. We also calculate the Fourier transform of the data to verify the spectral content of the signal and find a single large peak at the expected frequency. 

\section{Detuning noise model} \label{sec:detnoisemodel}

Since our device is fabricated with enriched \ce{^{28}Si}, the fluctuations in the ``magnetic'' control axis $A/2$ are expected to be small. Other work in ST qubits has shown that a dominant mechanism limiting the coherence is noise in exchange $J(\eps)$ induced by quasi-static noise on the detuning $\eps \rightarrow \eps+\eta$ \cite{dial2013,bertrand2015}, i.e.\ ``charge'' noise. Given a quasi-static noise on the detuning $\eta$ having zero mean and standard deviation $\sigma_\eps$, an ensemble average leads to a Gaussian decay of the coherence of the form
\ma{
	C(t) 
	&= \int_{-\infty}^{\infty} \D\eta P(\eta) \cos\offrac{t\Delta(\eps+\eta)}{\hbar} \\
	&= \Exp{-\offrac{t}{T_2^*}^2} \cos\offrac{t\Delta(\eps)}{\hbar}   ,
}
where $P(\eta) = \E{-\eta^2/2\sigma_\eps^2}/\sqrt{2\pi}\sigma_\eps$, $\Delta(\eps)$ is the energy gap $\Delta(\eps) = \sqrt{J^2+(A/2)^2}$, and
\ma{
	T_2^* = \frac{\sqrt{2} \hbar}{\sigma_\eps \abs{\partial\Delta/\partial\eps}}   .
}
Since the values we report for $T_2^*$ pertain to an ensemble average of measurements over a timescale of hours, our estimated detuning noise strength includes the effects of a secular drift component as well. While sufficiently large variations of the detuning can lead to Stark shifting of the contact hyperfine strength $A$, this Stark shifting effect should be small compared to the $\sigma_\eps \sim 9 \ueV$ that we observe.

\end{document}